\input{epsf}

\newcommand{\BOX}{\hbox {$\sqcap$ \kern -1em $\sqcup$}}

\newcounter{letter}
\newenvironment{alphalist}{
\begin{list}{{\normalshape(\alph{letter})}}{\usecounter{letter}}
}{\end{list}}

\newcommand{\g}{{\frak g}}
\newcommand{\so}{{\frak so}}

\newcommand{\R}{{\Bbb R}}
\newcommand{\C}{{\Bbb C}}

\renewcommand{\to}{\rightarrow}
\newcommand{\tensor}{\otimes}
\newcommand{\maps}{\colon}

\newcommand{\iso}{\cong}
\newcommand{\we}{\wedge}

\renewcommand{\H}{{\cal H}}

\newcommand{\T}{{\cal T}}

\newcommand{\SU}{{\rm SU}}

\newcommand{\SO}{{\rm SO}}

\newcommand{\Inv}{{\rm Inv}}

\newcommand{\xd}{{\rm d}}

        \newcommand{\be}{\begin{equation}}
        \newcommand{\ee}{\end{equation}}
        \newcommand{\ba}{\begin{eqnarray}}
        \newcommand{\ea}{\end{eqnarray}}
        \newcommand{\ban}{\begin{eqnarray*}}
        \newcommand{\ean}{\end{eqnarray*}}
        \newcommand{\barr}{\begin{array}}
        \newcommand{\earr}{\end{array}}

\documentstyle[12pt,amsfonts]{article}

\newcommand{\doubleY}[5] 
{
 { \raise6pt\hbox{$#1$}\atop 
\lower6pt\hbox{$#2$}}
  { \hbox to20pt{$\hfill #5\hfill $}\overwithdelims>< {} }
 { \raise6pt\hbox{$#3$}\atop 
\lower6pt\hbox{$#4$}}
 }
\newcommand{\singleY}[3] 
{
 { \raise6pt\hbox{$#1$}\atop 
\lower6pt\hbox{$#2$}}
  { \hbox to15pt{$\hfill$}\overwithdelims>. } \,#3\, 
 }

\begin{document}

      \begin{center}
      {\bf The Quantum Tetrahedron in 3 and 4 Dimensions \\}
      \vspace{0.5cm}
      {\em John C.\ Baez \\}
      \vspace{0.3cm}
      {\small Department of Mathematics, University of California\\
      Riverside, California 92521 \\
      USA\\ }
      \vspace{0.5cm}
      {\em John W.\ Barrett\\}
      \vspace{0.3 cm}
      {\small Department of Mathematics, University of Nottingham \\
       University Park, Nottingham, NG7 2RD \\
       United Kingdom \\}
      \vspace{0.3cm}
      {\small email: baez@math.ucr.edu, jwb@maths.nott.ac.uk \\}
      \vspace{0.3cm}
      {\small March 15, 1999 \\ }
      \end{center}

\begin{abstract} 
Recent work on state sum models of quantum gravity in 3 and 4 dimensions
has led to interest in the `quantum tetrahedron'.  Starting with a
classical phase space whose points correspond to geometries of the
tetrahedron in $\R^3$, we use geometric quantization to obtain a Hilbert
space of states.  This Hilbert space has a basis of states labeled by
the areas of the faces of the tetrahedron together with one more quantum
number, e.g.\ the area of one of the parallelograms formed by midpoints
of the tetrahedron's edges.  Repeating the procedure for the tetrahedron
in $\R^4$, we obtain a Hilbert space with a basis labelled solely by the
areas of the tetrahedron's faces.  An analysis of this result yields a
geometrical explanation of the otherwise puzzling fact that the
quantum tetrahedron has more degrees of freedom in 3 dimensions than
in 4 dimensions.  
\end{abstract}

\section*{Introduction} 

State sum models for quantum field theories are constructed by giving
amplitudes for the simplexes in a triangulated manifold. The simplexes
are labelled with data from some discrete set, and the amplitudes depend
on this labelling.  The amplitudes are then summed over this set of
labellings, to give a discrete version of a path integral. When the
discrete set is a finite set, then the sum always exists, and so this
procedure provides a {\it bona fide} definition of the path integral.

State sum models for quantum gravity have been proposed based on the Lie
algebra $\so(3)$ and its $q$-deformation.  Part of the labelling scheme
is then to assign irreducible representations of this Lie algebra to
simplexes of the appropriate dimension.  Using the $q$-deformation, the
set of irreducible representations becomes finite.  However, we will
consider the undeformed case here as the geometry is more elementary.

Irreducible representations of $\so(3)$ are indexed by a non-negative
half-integers $j$ called spins.  The spins have different
interpretations in different models.  In the Ponzano-Regge model of
3-dimensional quantum gravity \cite{PR}, spins label the edges of a
triangulated 3-manifold, and are interpreted as the quantized lengths of
these edges.  In the Ooguri-Crane-Yetter state sum model \cite{CY,O},
spins label triangles of a triangulated 4-manifold, and the spin is
interpreted as the norm of a component of the $B$-field in a $BF$ 
Lagrangian \cite{B}.  There is also a state sum model of 4-dimensional 
quantum gravity in which spins label triangles \cite{BC}.  Here the spins 
are interpreted as areas \cite{BC}.  

Many of these constructions have a topologically dual formulation.  The
dual 1-skeleton of a triangulated surface is a trivalent graph, each of
whose edges intersect exactly one edge in the original triangulation. 
The spin labels can be thought of as labelling the edges of this graph, 
thus defining a spin network \cite{B2,B3,Penrose,RS}.  In the
Ponzano-Regge model, transition amplitudes between spin networks can
be computed as a sum over labellings of faces of the dual 2-skeleton
of a triangulated 3-manifold \cite{HP,TV}.  Formulated this way,
we call the theory a `spin foam model'.

A similar dual picture exists for 4-dimensional quantum gravity.  The dual 
1-skeleton of a triangulated 3-manifold is a 4-valent graph each of
whose edges intersect one triangle in the original triangulation 
\cite{Markopoulou}.  The labels on the triangles in the 3-manifold can 
thus be thought of as labelling the edges of this graph.  The graph is 
then called a `relativistic spin network'.  Transition amplitudes
between relativistic spin networks can be computed using a spin foam 
model.  The path integral is then a sum over labellings of faces of a 
2-complex interpolating between two relativistic spin networks \cite{B4}.

In this paper we consider the nature of the quantized geometry of a
tetrahedron which occurs in some of these models, and its relation to
the phase space of geometries of a classical tetrahedron in 3 or 4
dimensions.  Our main goal is to solve the following puzzle: why does
the quantum tetrahedron have fewer degrees of freedom in 4 dimensions
than in 3 dimensions?  This seeming paradox turns out to have a simple
explanation in terms of geometric quantization. The picture we develop 
is that the four face areas of a quantum tetrahedron in four dimensions 
can be freely specified, but that the remaining parameters cannot, due
to the uncertainty principle.  

In the rest of this section we briefly review the role of the triangle
in 3-dimensional quantum gravity and that of the tetrahedron in
4-dimensional quantum gravity.  We also sketch the above puzzle and its
solution.  In the following sections we give a more detailed treatment.

\subsection{The triangle in 3d gravity}

In the Ponzano-Regge model, states are specified by labelling the edges
of a triangulated surface by irreducible representations of $\so(3)$.
Using ideas from geometric quantization, the spin-$j$ representation can
be thought of as the Hilbert space of states of a quantized vector of
length $j$.  Once spins $j_1,j_2,j_3$ for the edges of a triangle are
specified, a quantum state of the geometry of the triangle is an element
\[    \psi \in j_1 \otimes j_2 \otimes j_3. \]
We require that this element satisfy a quantized form of the condition
that the three edge vectors for a triangle sum to zero.  The
quantization of this condition is that $\psi$ be invariant under the
action of $\so(3)$.  As long as the spins satisfy the Euclidean triangle
inequalities, there exists a unique element with this  property, up to a
constant factor\footnote{The normalisation of this element is fixed by
the inner products on $j_1$, $j_2$ and $j_3$.  This fixes a canonical
element up to a phase.  Its phase is fixed by a coherent set of
conventions for planar diagrams of spin networks \cite{Moussouris}.}. 
This reflects the fact that the geometry of a Euclidean triangle is
specified entirely by its edge lengths.  This unique invariant element
is denoted graphically by the spin network
\[  \singleY {j_1} {j_2} {j_3}. \]
We call it a {\it vertex} because it corresponds to a vertex of 
the spin network dual to the triangulation.  

The uniqueness of the vertex can be explained in terms of geometric
quantization.  This explanation serves as the prototype of the discussion
below for the uniqueness of the vertex in the
4-dimensional model.  Any irreducible representation of $\so(3)$ can be
realised by an action of $\SU(2)$ on the space of holomorphic sections
of some line bundle over $S^2$.  Elements of the space $j_1 \tensor j_2
\tensor j_3$ thus correspond to holomorphic sections of the tensor
product line bundle on $S^2\times S^2\times S^2$.  A vertex therefore
corresponds to an $\SU(2)$-invariant section of this tensor product line
bundle.

In general, invariant holomorphic sections of a line bundle over a
K\"ahler manifold correspond to sections of a line bundle over the
symplectic reduction of this manifold by the group action \cite{GS2}.
Moreover, symplectic reduction eliminates degrees of freedom in
conjugate pairs.  Starting from the 6-dimensional manifold $S^2\times
S^2\times S^2$ and reducing by the action of the 3-dimensional group
$\SU(2)$, we are thus left with a 0-dimensional reduced space, since
$6-3-3 = 0$.  

In fact, when the spins $j_i$ satisfy the triangle inequality, the
reduced space is just a single point.  To see this, think of a
point in $S^2\times S^2\times S^2$ as a triple of bivectors
with lengths equal to $j_1,j_2,j_3$, respectively.  The 
constraint generating the $\SU(2)$ action is that these bivectors 
sum to zero.  When this constraint holds, the bivectors form a 
triangle in $\so(3)^*$ with edge lengths $j_1,j_2,j_3$.  Since 
all such triangles are contained in one orbit of $\SU(2)$, the 
reduced phase space is a single point.  This explains why the space 
of vertices is 1-dimensional: any vertex corresponds to a section 
of a line bundle over this point.

\subsection{The tetrahedron in 4d gravity}

A modification of the Ponzano-Regge model has been proposed to give a
4-dimensional state sum model of quantum gravity \cite{BC}.  In these
models states are described by labelling the triangles of a triangulated
3-manifold by irreducible representations of $\so(4)$.  However, not all
irreducible representations are allowed as labels, only certain ones
called `balanced' representations.  This restriction can understood as
follows.

In the Ponzano-Regge model, the basic idea was to describe triangles by
labelling their edges with vectors in $\R^3$, and then to quantize this
description.  Similarly, in the 4-dimensional model we describe
tetrahedra by labelling their faces with bivectors in $\R^4$, and then we
quantize this description.  A {\it bivector} in $\R^4$ is just an element of
the second exterior power $\Lambda^2 \R^4$.  Given an oriented triangle
in $\R^4$ we may associate to it the bivector formed as the wedge
product of any two of its edges taken in cyclic order.   

Irreducible representations of $\so(4)\cong\so(3)\oplus\so(3)$ are
indexed by pairs of spins $(j,k)$.  The $(j,k)$ representation can be
thought of as the Hilbert space of states of a quantized bivector in 4
dimensions whose self-dual and anti-self-dual parts have norms $j$ and
$k$, respectively.  Classically, a bivector in $\R^4$ comes from
a triangle as above if and only if its self-dual and anti-self-dual
parts have the same norm.  We impose this restriction at the quantum
level by labelling triangles only with representations for which $j = k$.
These are called {\it balanced} representations of $\so(4)$.
When a triangle is labelled by the balanced representation $(j,j)$,
the spin $j$ specifies its area.  Thus in going from three to
four dimensions, the geometric interpretation of a spin label changes
from the length of an edge to the area of a triangle.  

Having labelled its faces by balanced representations, the quantum state 
of the geometry of a tetrahedron in four dimensions is given by a vector
\[    \psi\in (j_1,j_1)\otimes(j_2,j_2)\otimes(j_3,j_3)\otimes(j_4,j_4). \]
As in the Ponzano-Regge model, we impose some conditions on this vector,
motivated by the geometry of the situation.  First, we require that this
vector be invariant under the action of $\so(4)$.  This is a quantization
of the condition that the bivectors associated to the faces of a tetrahedron
must sum to zero.  Second, we require that given any pair of faces,
e.g.\ 1 and 2, then in the decomposition
\[ (j_1,j_1)\otimes(j_2,j_2)\cong \bigoplus_{j,k} (j,k) \]
the components of $\psi$ are zero except in the balanced summands,
$(j,j)$.  There are three independent conditions of this sort, one for
the pair 1--2, one for the pair 1--3 and one for the pair 1--4.  (Since
the vector is invariant, the summands of the pair 3--4 which occur are
exactly the same as for 1--2, and so on, so we do not need consider the
other three pairs.)  This quantizes the condition that any pair of faces of
a tetrahedron must meet on an edge, geometrically a rather strong condition 
as a generic pair of planes in $\R^4$ meet at a point, not a line.
Vectors $\psi$ meeting all these requirements are called {\it vertices}, 
now corresponding to the vertices of the 4-valent relativistic spin network 
dual to the triangulation of a 3-manifold.

Taking into account these conditions, the vertex is unique up to a
constant factor.  The formula for the vertex was given by Barrett and
Crane \cite{BC}, and an argument for its uniqueness was first given by
Barbieri \cite{Barb2}.  This argument depended on an assumption that
some associated 6$j$-symbols do not have too many `accidental zeroes'.  A
simpler argument relying on the same assumption is given here.  A proof
of the uniqueness without any assumptions has been given by Reisenberger
\cite{R2}.  

A basis for the invariant vectors is given by 
\[ B^{12}_{jk}= \doubleY{j_1}{j_2}{j_3}{j_4}j \bigotimes 
\doubleY{j_1}{j_2}{j_3}{j_4}k. \]
Similar bases $B^{13}_{jk}$ and $B^{14}_{jk}$ are defined using the spin 
networks which couple the pairs 1--3 or 1--4 instead of 1--2. Any 
solution of the constraints for the quantum tetrahedron must be
\be 
\label{one} 
\psi=\sum_j \lambda_j B^{12}_{jj}=
\sum_k \mu_k B^{13}_{kk}=
\sum_l \nu_l B^{14}_{ll}.
\ee
In other words, $\psi$ is a symmetric rank two tensor on the invariant
subspace $\Inv(j_1 \tensor j_2 \tensor j_3 \tensor j_4)$ 
which is diagonal in three different bases.  The change of basis matrix 
is, for each pair, a 6$j$-symbol. Since the 6$j$-symbol satisfies an 
orthogonality relation, there is a canonical solution to these 
constraints, given by taking $\psi$ to be (inverse of the) inner product 
in the orthogonality relation, $\lambda_j=(-1)^{2j}\dim j$. This gives 
the vertex described by Barrett and Crane \cite{BC}.

Now suppose there is another solution $\lambda'_j$. Then
$\lambda'_j/\lambda_j$ are the eigenvalues of a linear operator which is
diagonal in each basis. It follows that the eigenspaces for each
distinct eigenvalue are preserved under the change of basis, so the
6$j$-symbol must be zero for each choice of spin $j$ corresponding to
one eigenvalue in $B^{12}_{jj}$ and spin $k$ corresponding to a
different eigenvalue in $B^{13}_{kk}$.  Assuming the 6$j$-symbol does not
have accidental zeroes in this way, there cannot be two distinct
eigenvalues, so the two solutions are just proportional to each
other.  Since accidental zeroes of the 6$j$ symbols are rather sparse
\cite{BL}, we expect the vertex to be unique --- as was indeed shown
by Reisenberger.  

Still, from a geometrical viewpoint the uniqueness of the vertex is
puzzling.  After all, labelling a triangle with a balanced
representation of $\so(4)$ only specifies the area of the triangle.
Fixing the values of the areas of four triangles does not specify the
geometry of a tetrahedron uniquely.  The geometry of a tetrahedron is
determined by its six edge lengths.  Since the areas of the faces are
only four parameters, there will be typically a 2-dimensional moduli
space of tetrahedra with given face areas.  Why then is a state of
the quantum tetrahedron in four dimensions uniquely determined by the
areas of its faces?

This puzzle is resolved here by developing a deeper understanding
of the constraints, using geometric quantization.  Essentially the
issue is to understand the difference between a tetrahedron
embedded in $\R^3$ and a tetrahedron embedded in $\R^4$.  In classical
geometry, these are not too different.  In both cases we can describe
the tetrahedron using bivectors for faces that sum to zero.  But in 
the 4-dimensional situation there are also extra constraints implying that
all four faces lie in a common hyperplane.  When these are satisfied we are 
essentially back in the 3-dimensional situation.

Quantum mechanically, however, these extra constraints dramatically
reduce the number of degrees of freedom for the tetrahedron.  This is a
standard phenomenon in quantum theory.  A set of constraints
introduced by operator equations
\[ \widehat{H}_i \psi =0,\quad\quad i=1,\ldots, n \]
implies automatically that the quantum state vector $\psi$ is invariant
under the group of symmetries generated by the set of operators
$\widehat{H}_i$.  The classical counterpart is symplectic
reduction, or more generally Poisson reduction \cite{MR}, in which one
restricts the phase space to the subspace given by the constraints
$H_i=0$, and then passes to the quotient space of orbits of the symmetry
group generated by the constraints $H_i$.  

Our explanation of the uniqueness of the vertex for the quantum 
tetrahedron in four dimensions is as follows.   The Hilbert space 
\[  (j_1,j_1)\otimes(j_2,j_2)\otimes(j_3,j_3)\otimes(j_4,j_4) \] 
corresponds to the classical phase space consisting of 4-tuples of
simple bivectors having norms given by the four spins $j_i$.  This phase
space is 16-dimensional.  The subspace of quantum states invariant under
$\so(4)$ corresponds to the classical phase space of four such bivectors
summing to zero, considered modulo the action of the rotation group
$\SO(4)$ on all four bivectors simultaneously.  This phase space is a
4-dimensional symplectic manifold (since $16-6-6 = 4$).  The constraints
(\ref{one}) which force the triangles to intersect on edges are two
independent equations.  Symplectic reduction with two constraints
reduces the 4-dimensional manifold to a 0-dimensional one (since $4-2-2 = 0$),
in fact just a single point.  This explains why the space of vertices 
is 1-dimensional.  

This argument can also be phrased in terms of the uncertainty principle.
There are two constraint equations $H_i = 0$ that force the four faces of the
tetrahedron to lie in a common hyperplane. These
variables $H_i$ are canonically conjugate to the two variables
determining the shape of a tetrahedron with faces of fixed area.
Therefore by the uncertainty principle, the shape of the tetrahedron is
maximally undetermined.

\section{Quantum bivectors} \label{bivect}

By a {\it bivector} in $n$ dimensions we mean an element of $\Lambda^2
\R^n$.  A bivector records some of the information of the geometry of a
triangle in $\R^n$.  An oriented triangle has three edge vectors, $e,f,g$
which cycle around the triangle in that order. Then the bivector for the
triangle is  
\[     E = e\wedge f = f\wedge g = g\wedge e. \]
These are equal due to the equation $e+f+g=0$. Therefore $b$ depends
only on the triangle and its orientation, not on a choice of 
edges.  The bivector determines a 2-dimensional plane in $\R^n$ with an
orientation, and the norm of the bivector is twice the area of the
triangle.  However, no further details of the geometry of the triangle
are recorded.

A bivector formed as the wedge product of two vectors is called {\it
simple}.   By the above remarks, we may think of any simple bivector 
as an equivalence class of triangles in $\R^n$.  This interpretation
becomes important in the next section, where we describe tetrahedra 
in terms of bivectors.  

In three dimensions or below every bivector is simple, but this is not
true in higher dimensions.  For example, if $a_1, a_2, a_3, a_4$ are
linearly independent, then $a_1\wedge a_2+a_3\wedge a_4$ is not a simple
bivector.   In general, a bivector $E$ is simple if and only if $E\wedge
E=0$.  In four dimensions another criterion is also useful.  The
Euclidean metric and standard orientation on $\R^4$ determine a Hodge
star operator
\[       \ast \maps \Lambda^2 \R^4 \to \Lambda^2 \R^4.  \]
Since $\ast^2 = 1$, we may decompose any bivector into its
self-dual and anti-self-dual parts:
\[        E = E^+ + E^- , \qquad \ast E^{\pm} = \pm E^\pm .\]
The bivector is simple if and only if its self-dual and anti-self-dual
parts have the same norm.

Using the Euclidean metric $\eta$ on $\R^n$ we may identify 
bivectors with elements of $\so(n)^*$.  Explicitly, this isomorphism
$\beta\maps \Lambda^2 \R^n \to \so(n)^\ast$ is given by 
\[ \beta(e \we f)(l)  = \eta(le,f)  \]
for any bivector $e \we f$ and any $l \in \so(n)$.   As recalled below, the 
dual of any Lie algebra has a natural Poisson structure.  This allows us 
to treat the space of bivectors as a classical phase space.  Using geometric 
quantization, we can quantize this phase space and construct a Hilbert 
space which we call the space of states of a `quantum bivector'.  While this
construction works in any dimension, we initially concentrate on
dimension three.  Then we turn to dimension four.  In this case, the
Hilbert space we construct has a subspace representing the states 
of a `simple' quantum bivector.  

\subsection{Kirillov-Kostant Poisson structure}  \label{bivect.poisson}

As shown by Kirillov and Kostant \cite{Kirillov,Kostant}, the
dual of any Lie algebra $\g$ is a vector space with an additional
structure that makes it a Poisson manifold.  This is a manifold with a
Poisson bracket on its algebra of functions.  We are mainly interested
in the Lie algebra $\so(3)$ here, but $\so(4)$ is also important to us,
so we briefly recall the general construction.  

Elements $l,m$ of the Lie algebra determine linear functions on the dual
vector space $\g^*$.  The Lie bracket $[l,m]$ can likewise be thought of
as a function on $\g^*$.  This defines a Poisson bracket on the linear
functions, and this extends uniquely to determine a Poisson bracket on
the algebra of all smooth functions.  Indeed, there is a bivector field
$\Omega$ on the manifold $\g^*$ such that
\[ [l,m]=\Omega(\xd l,\xd m) \]
Then the Poisson bracket in general is
\[  \{f,g\}=\Omega(\xd f,\xd g). \]
Since this formula is linear in $\xd g$, it is given by the evaluation of $\xd
g$ on a vector field $f^\#$ on $\g^*$
\[  \{f,g\}=\xd g(f^\#)=f^\#g. \]
In particular, each Lie algebra element $l$ determines a vector field $l^\#$.
If $x\in\g^*$, then the value of $l^\# m$ at $x$ is
\[ (l^\# m)(x) =\langle x,[l,m]\rangle. \]
Thus if $\g$ is the Lie algebra of a Lie group $G$, the vector fields
$l^\#$ are given by the natural action of $G$ on $\g^*$, the coadjoint action.

Coordinate formulae are sometimes illuminating for the above.  Pick a
basis $x^i$ for $\g$, which gives coordinate functions on $\g^*$.  Then
$[l,m]=[l_i x^i,m_j x^j]=l_i m_j c^{ij}_k x^k$, where $c^{ij}_k$ are the
structure constants of the Lie algebra.  Then $\Omega^{ij}=c^{ij}_k x^k$,
and $l_\#^j =l_i\Omega^{ij}$.

A 2-form $\omega$ is compatible with the Poisson structure if
$$\omega(l^\#,m^\#)=\{l,m\}.$$  This only determines $\omega$ on the
span of the $l^\#$, namely the tangent space to the coadjoint orbit.
Therefore, $\omega$ is defined as a 2-form on each orbit, called a
symplectic leaf of the Poisson manifold.  The Poisson bivector $\Omega$
is tangent to each leaf and non-degenerate as a bilinear form on the
cotangent bundle of each leaf.  The symplectic form $\omega$ is the
inverse of this.

In the case of $\so(3)$, the Lie algebra can be identified with $\R^3$
with its standard vector cross product, the dual space can also be
identified with $\R^3$, and then dual pairing
$\langle\cdot,\cdot\rangle$ becomes the Euclidean inner product. Then
\[ \{l,m\}(x) = \langle x,[l,m]\rangle      \]
which looks the same as before, but now the right-hand side is the triple
scalar product of vectors in $\R^3$.  The 2-form is
\[ \omega(a,b)={1\over x^2}\langle x,[a,b]\rangle.  \]

In this case, the coadjoint action is the familiar action of rotations
on the vector space of angular momenta, and so the symplectic leaves are
spheres centered at the origin.  The integral of $\omega$ over one of
these spheres of Euclidean radius $r$ is $\pm 4\pi r$, the sign depending on
the chosen orientation of the sphere.  This differs from the Euclidean area
due to the scaling factors in the formulae.

\subsection{Quantization} \label{bivect.quant}

Throughout this paper we quantize Poisson manifolds by constructing a
Hilbert space for each symplectic leaf and then forming the direct sum
of all these Hilbert spaces.  We construct the Hilbert space for each
leaf using geometric quantization, or more precisely, K\"ahler
quantization \cite{GS,Sniatycki,Woodhouse}.  For this we need to
introduce some extra structure on each leaf.  First we choose a complex
structure $J$ on the leaf that preserves the symplectic form $\omega$,
making the leaf into a K\"ahler manifold.  Then we choose a holomorphic
complex line bundle $L$ over the leaf, called the {\it prequantum line
bundle} equipped with a connection whose curvature equals $\omega$:
\[ \omega(\xi,\eta)s=i\left( \nabla_\xi\nabla_\eta - \nabla_\eta  \nabla_\xi
- \nabla_{[\xi,\eta]}\right)s. \]
For this, the symplectic leaf must be {\it integral}: the closed 2-form
$\omega/2\pi$ must define an integral cohomology class.  In this case,
we define the Hilbert space for the leaf to be the space of
square-integrable holomorphic sections of $L$.  For nonintegral leaves,
we define the Hilbert space to be 0-dimensional.

In the case of $\so(3)^*$, the integral symplectic leaves are the
spheres centered at the origin for which the integral of $\omega$ is
$2\pi$ times an integer.  These are spheres $S_j$ with radii given by
nonnegative half-integers $j$ --- honest 2-spheres for $j > 0$, and
a single point for $j = 0$.  Each sphere $S_j$ with $j > 0$ has a
complex structure $J$ corresponding to the usual complex structure on
the Riemann sphere.  Explicitly, using the identification of
$\so(3)^\ast$ and $\so(3)$, the formula for $J$ at a point $x$
\[  J(a) = {1\over |x|} [x,a] .\]
In other words, $J$ rotates tangent vectors a quarter turn
counter-clockwise.  Clearly $J$ is a complex structure preserving the
symplectic form $\omega$.  The sphere thus becomes a K\"ahler
manifold with Riemannian metric given by
\[  g(a,b)=\omega(a,Jb)={1\over |x|}\langle a,b\rangle. \]
Like the symplectic structure itself, this differs from the standard
induced metric by a scale factor.

When $j = 0$, $S_j$ is trivially a K\"ahler manifold.  In this case the
prequantum line bundle is trivial and the Hilbert space of holomorphic
sections is 1-dimensional.  When $j = {1\over 2}$, we may choose the
prequantum line bundle to be the spinor bundle over $S_j$.  The Hilbert
space of holomorphic sections is then isomorphic to $\C^2$.  For other
values of $j$ we choose $L$ to be the $2j$th tensor power of the spinor
bundle over $S_j$.  The Hilbert space of holomorphic sections is then
isomorphic to the $2j$th symmetrized tensor power $S^n \C^2$, which is
$(2j + 1)$-dimensional.  We may associate to any Lie algebra element
$l \in \so(3)$ a self-adjoint operator $\hat l$ on this Hilbert space,
given by
\[
 \hat l\colon s \mapsto i\nabla_{l^\#}s - ls  
\]
On the right hand side, $l$ is regarded as a function on the coadjoint
orbit, and multiplies the section $s$ pointwise.  This formula gives a
representation of $\so(3)$ which is just the usual spin-$j$ representation.  

Taking the sum over all symplectic leaves, we obtain the {\it
Hilbert space of a quantum bivector in 3 dimensions},
\[   \H = \bigoplus_j j . \]
This is the basic building-block of all the more complicated quantized
geometrical structures appearing in state sum models of quantum gravity
\cite{B4}.  Let $E^i$ be a basis of $\so(3)$ satisfying the 
Poisson bracket relations
\[       \{ E^1, E^2\} = E^3,\qquad
         \{ E^2, E^3\} = E^1,\qquad
         \{ E^3, E^1\} = E^2  .\]
when thought of as coordinate functions on $\so(3)^\ast$.  Then by the
above we obtain self-adjoint operators $\hat E^i$ on $\H$ satisfying 
the usual angular momentum commutation relations: 
\[       [\hat E^1,\hat E^2] = i\hat E^3,\qquad
         [\hat E^2,\hat E^3] = i\hat E^1,\qquad
         [\hat E^3,\hat E^1] = i\hat E^2  .\]
We think of these operators as observables measuring the 3 components of
the quantum bivector.   This interpretation is justified by the fact that
these operators can also be obtained by geometrically quantizing the
3 coordinate functions on the space of bivectors.  Their failure to commute
means that the components of a quantum bivector cannot in general be
measured simultaneously with complete precision.

\subsection{The 4-dimensional case} \label{bivect.4d}

In any dimension, starting from the Kirillov-Kostant Poisson structure
on the dual of $\so(n)$, one can use geometric quantization to construct
a Hilbert space describing the states of a quantum bivector.  By the
Bott-Borel-Weil theorem \cite{Bott}, this Hilbert space always turns out
to be the direct sum of all the irreducible unitary representations of
$\so(n)$.  The 4-dimensional case is particularly simple, since
we can reduce it to the previously treated 3-dimensional case using the
isomorphism $\so(4) \iso \so(3) \oplus \so(3)$.  This isomorphism
corresponds to the splitting of a bivector into its self-dual and
anti-self-dual parts.     

Using this isomorphism, it follows that $\so(4)^\ast$ with its
Kirillov-Kostant Poisson structure is the product of two copies of the
Poisson manifold $\so(3)^\ast$.  In particular, any symplectic leaf is
of the form $S_j \times S_k$, that is, the product of a sphere of radius
$j$ and a sphere of radius $k$, where $j,k$ are independent arbitrary
spins.  We can make this symplectic leaf into a K\"ahler manifold by
taking the product of the previously described K\"ahler structures on
$S_j$ and $S_k$.  Similarly, we can equip the leaf with a line bundle
given by the tensor product of the previously  described line bundles
over $S_j$ and $S_k$.  As a result, when we geometrically quantize the
leaf, we obtain a Hilbert space equal to the tensor product of the
spin-$j$ representation of the `left-handed' copy of $\so(3)$ and the
spin-$k$ representation of the `right-handed' copy.  Taking the
direct sum over all leaves, we thus obtain the Hilbert space
\[   \H \tensor \H  = \bigoplus_{j,k} j \tensor k  .\]
This is just the direct sum of all irreducible representations
$(j,k)$ of $\so(4)$.  

However, $\so(4)^\ast$ has another Poisson structure, obtained by
reversing the sign of the Poisson structure on the anti-self-dual
summand in $\so(3)^\ast \oplus \so(3)^\ast$.  We call this the {\it
flipped} Poisson structure.  It turns out that the flipped Poisson
structure is the right one for our purposes, as with this Poisson 
structure the quantum theory determines the {\it chirality} of the 
tetrahedron correctly, as is discussed in Section \ref{4dtet.quant}. The 
issue is to use the correct Poisson structure on $\Lambda^2 \R^4$, which 
we identified with $\so(4)^*$ using the isomorphism $\beta$ described at
the beginning of this section.  We could achieve the same 
Poisson structure on $\Lambda^2 \R^4$ by starting with the standard 
Poisson structure on $\so(4)^*$ and instead using the isomophism 
$\beta \circ \ast$, which differs by the Hodge dual $\ast$ acting on 
the bivectors.  We return to this issue in Section \ref{considerations}.

With its flipped Poisson structure, $\so(4)^\ast$ again has integral
symplectic leaves of the form $S_j \times S_k$; the only 
difference now is that the sign of the symplectic structure is reversed
on $S_k$.  We can thus make $S_k$ into a K\"ahler manifold with the 
Riemannian metric of the previous section but with the opposite complex
structure.  Proceeding as before, but with this modification, we can
geometrically quantize $\so(4)^\ast$ and obtain the {\it Hilbert space
of a quantum bivector in 4 dimensions}
\[           \H \tensor \H^* \iso \bigoplus_{j,k} j \tensor k^* \]
where starring stands for taking the complex conjugate Hilbert
space, which is canonically isomorphic to the dual.  

Using the splitting of $\so(4)^*$ into self-dual and anti-self-dual
copies of $\so(3)^*$, we can put coordinate functions $E^{+i}, E^{-i}$ 
on $\so(4)^*$ whose Poisson brackets with respect to the flipped
Poisson structure are:
\[       \{E^{+i}, E^{+j} \}   = \epsilon^{ijk} E^{+k}  , \qquad
           \{E^{-i}, E^{-j} \} = -\epsilon^{ijk} E^{-k}  , \qquad
           \{E^{+i}, E^{-j} \} =  0 . \]
The minus sign is the result of using the flipped Poisson structure.
Using the recipe discussed in the previous section, we thus obtain
self-adjoint operators $\hat E^{\pm i}$ on $\H \tensor \H^*$ satisfying
the following commutation relations:
\[  [{\hat E}^{+i}, {\hat E}^{+j} ] = i\epsilon^{ijk} {\hat E}^{+k} , \qquad
      [{\hat E}^{-i}, {\hat E}^{-j} ] = -i\epsilon^{ijk} {\hat E}^{-k} , \qquad
      [{\hat E}^{+i}, {\hat E}^{-j} ] =  0 .  \]

As already noted, a bivector $b$ in 4 dimensions is simple if
and only if its self-dual and anti-self-dual parts have the same norm.
We may impose this constraint at the quantum level, obtaining the
subspace of $\H \tensor \H^*$ consisting of states $\psi$ for which the 
Casimir operators
\[ K^\pm  =  \sum_{k=1}^3 (\hat E^{\pm k})^2 \]
are equal:
\[      K^+ \psi = K^- \psi .\]
Alternatively, we may impose the constraint classically and then use 
geometric quantization to obtain a Hilbert space from the resulting Poisson
manifold.  Either way, we obtain the {\it Hilbert space of a simple
quantum bivector in 4 dimensions},
\[         \bigoplus_j j \tensor j^\ast \;\subset\; \H \tensor \H^* . \]

\section{The Quantum Tetrahedron in 3 Dimensions} \label{3dtet}

A tetrahedron in $\R^n$ with labelled vertices, modulo translations, is
determined by three vectors $e_1, e_2, e_3$, which are the edge vectors
for three edges pointing out from a common vertex.  The three triangular
faces of the tetrahedron meeting at this vertex have bivectors
 \[       E_1 = e_2 \we e_3, \; E_2 = e_3 \we e_1, \; E_3 = e_1 \we e_2  ,\]
while the bivector for the fourth face is
\[ E_4 = -E_1 - E_2 - E_3. \]
If the vectors $e_i$ are linearly dependent, then the $E$'s are
all multiples of each other or zero.   But in dimensions three or more, 
the vectors are generically linearly independent.  From now on, we only 
consider this generic situation.  

The map from triples of vectors to triples of bivectors is generically
two-to-one, because if all three vectors are replaced by their
negatives, then the bivectors are unchanged.  Moreover, when the vectors
$e_i$ are linearly independent, this operation is the only operation
possible that changes the vectors but not the corresponding bivectors.

In dimension three, the map is also generically `half-onto', in the
sense that all positively oriented triples of bivectors $E_1,E_2,E_3$
arise, but never negatively-oriented ones.  In three dimensions
$\Lambda^2 \R^3\cong \R^3$. With this identification the $\wedge$
operation is the vector cross product $\times$, so one can compute that
\[ E_1 \cdot (E_2 \times E_3) = V^2 \ge 0 \] 
where $V= e_1 \cdot (e_2 \times e_3)$ is 6 times the oriented volume of
the tetrahedron.   One also has
\[ E_1 \times E_2 = V e_3 \]
etc., so that from a positively oriented triple $E_1, E_2, E_3$ one can
compute the original $e$'s up to sign, i.e., obtaining either
$(e_1,e_2,e_3)$ or $(-e_1,-e_2,-e_3)$. The distinction between these two
triples of $e$'s is the orientation of the tetrahedron.  

To summarise, a non-degenerate tetrahedron in 3 dimensions can be
described by 4 bivectors, satisfying the {\it closure constraint}
\[ E_1 + E_2 + E_3 + E_4 = 0 \]
together with the {\it positivity constraint}
\[ E_1 \cdot (E_2 \times E_3) >0. \]
The positivity constraint
can be written in a number of equivalent ways by applying an
even permutation to the 4 numbers $1234$. For example, the permutation
$(14)(23)$ changes this formula to
\[ E_4 \cdot (E_3 \times E_2) >0 \]
which is equivalent, using the closure constraint.  Geometrically, the
different formulae correspond to calculating $V^2$ using different
vertices of the tetrahedron as the distinguished vertex.

\subsection{Poisson structure} \label{3dtet.poisson}

By the above we may describe a tetrahedron in 3 dimensions using a
4-tuple of bivectors $E_1,\dots,E_4$ satisfying the closure and
positivity constraints.   In this section we think of these bivectors as
elements of $\so(3)^\ast$.   To obtain a Poisson structure on the space
of geometries of a tetrahedron in 3 dimensions, we start by taking the
product of 4 copies of $\so(3)^\ast$ with its Kirillov-Kostant Poisson
structure, obtaining a Poisson structure on $(\so(3)^\ast)^4$.  Then we
perform Poisson reduction with respect to the closure constraint, as
follows.  

First we form the constraint submanifold
\[ C = \{ E_1 + E_2 + E_3 + E_4 = 0 \} \subset (\so(3)^\ast)^4 .\]
The closure constraint generates the diagonal $\SO(3)$ action on 
$(\so(3)^\ast)^4$, which preserves $C$.   To obtain the reduced space,
we take the quotient of $C$ by this $\SO(3)$ action.   We denote this
quotient by $T$, and call it the {\it phase space of a tetrahedron in 3
dimensions}.  Perhaps strictly speaking we should reserve this name for
the subset of $T$ where the positivity constraint holds.  However,
this smaller phase space is hard to quantize, so it is more convenient
to quantize all of $T$ and impose the positivity constraint at the
quantum level.  In fact, in applications to quantum gravity it
may be best not to impose the positivity constraint at all.  For more
discussion of this issue see Section \ref{considerations}.

\begin{figure}[h] 
\centerline{\epsfysize=1.5in\epsfbox{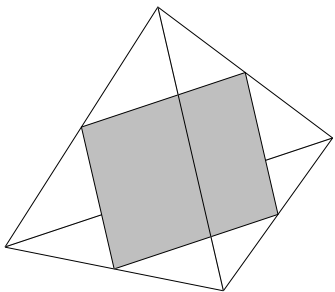}}
\medskip 
\caption{Parallelogram formed by midpoints of the tetrahedron's edges} 
\label{parallelogram}
\end{figure}

Any $\SO(3)$-invariant function on $(\so(3)^\ast)^4$ determines a
function on $T$.  We use this to define the following functions on  $T$:
\[   A_i = |E_i|,  \] 
\[   A_{ij} = |E_i + E_j|, \] 
and  
\[  U = E_1 \cdot (E_2 \times E_3). \] 
These quantities have nice geometrical
interpretations.  The positivity constraint is $U > 0$.  Thus if $U > 0$
there is a tetrahedron corresponding to the vectors $E_i$, while if $U <
0$ there is a tetrahedron corresponding to the vectors $-E_i$.  In
either case, $|U|$ is 36 times the square of the volume of this tetrahedron.  
The quantity $A_i$ is twice the area of the $i$th face of the tetrahedron. 
Similarly, the quantity $A_{ij}$ is 4 times the area of the
parallelogram with vertices given by the midpoints of the edges of the
tetrahedron that are contained in either the $i$th or $j$th face of the
tetrahedron, but not both.  (See Fig.\ \ref{parallelogram}.) 
Alternatively, $A_{ij}$ is equal to $|u\wedge v|$, where $u$ is the
displacement vector for the edge common to the $i$th and $j$th faces,
and $v$ the displacement vector for the edge common to the other two
faces.   

As noted in our earlier work, these quantities satisfy some relations 
\cite{B4}. In particular, if $ijkl$ is any permutation of $1234$ we have
\[     A_{ij} = A_{kl}  .\] 
We also have
\[     A_{12}^2 + A_{23}^2 + A_{31}^2 = A_1^2 + A_2^2 + A_3^2 + A_4^2 . \]

The map from $\SO(3)$-invariant functions on $(\so(3)^\ast)^4$
to functions on $T$ preserves Poisson brackets.  Since the functions
$|E_i|$ are constant on the symplectic leaves of $(\so(3)^\ast)^4$, 
it follows that the functions $A_i$ have 
vanishing Poisson brackets with all functions on $T$.  The functions
$A_{ij}$ do not.  In particular, for any even permutation
$ijkl$ of $1234$, we have
\[       \{A_{ij}^2, A_{ik}^2\} = 4 U. \]
When we quantize, this nonzero Poisson bracket leads to an uncertainty
relation between the areas of the different parallelograms formed by
midpoints of the tetrahedron's edges.  This is closely related to
the `noncommutativity of area operators' in loop quantum gravity
\cite{AL3}, which has similar classical origins.

To prepare for geometric quantization, let us study the symplectic
leaves of $T$.  These are obtained from the leaves of $(\so(3)^\ast)^4$
by symplectic reduction with respect to the closure constraint.    To
visualize the situation it is best to think of the $E_i$ as vectors in
$\R^3$.  The constraint submanifold $C$ then consists of configurations
of 4 vectors that close to form the sides of a (not necessarily planar)
quadrilateral in $\R^3$.   This  description in terms of quadrilaterals
allows us to apply the results of Kapovich, Millson \cite{KM},
Hausmann and Knutson \cite{HK}.
However, for the sake of a self-contained treatment we redo some of
their work.  Consider a symplectic leaf 
\[    \Lambda = \{ |E_i| = r_i \}  \]
in $(\so(3)^\ast)^4$.  Its intersection with $C$ consists of all
quadrilaterals with sides having fixed lengths  $r_1,\dots,r_4$.  The
corresponding leaf of $T$, namely $(\Lambda \cap C)/\SO(3)$, is thus the
space of such quadrilaterals modulo rotations.

Generically, the leaves of $T$ are 2-spheres.  In showing this, we 
shall exclude cases where $\SO(3)$ fails to act freely on $\Lambda \cap
C$.  These cases occur when $\Lambda \cap C$ contains a one-dimensional
configuration, i.e., one in which the $E_i$ are all proportional to one
another.  This happens when $r_1 + r_2 = r_3 + r_4$, $r_1 = r_2 + r_3 +
r_4$, or permutations of these.  We shall also exclude cases  where the
quotient $(\Lambda \cap C)/\SO(3)$ is a single point.   These cases
occur when one or more of the $r_i$ vanish, and also when $r_1 = r_2 +
r_3 + r_4$ or permutations thereof.  

\begin{figure}[h]
\centerline{\epsfysize=1.5in\epsfbox{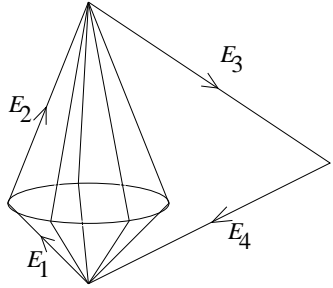}}
\medskip
\caption{Circle formed by rotating $E_1$ and $E_2$ about
$E_1 + E_2$}
\label{circle}
\end{figure}

Apart from these nongeneric cases, the leaf $(\Lambda \cap C)/\SO(3)$ is a
2-sphere.   To see this, note first that the level sets of the function
$A_{12}$ give a single point of this leaf at its maximum and minimum
values.  This is because at these points two of the $E_i$ are
proportional, so the quadrilateral they form reduces to a triangle in
$\R^3$.   The triangle is rigid when its edge lengths are specified, so
the set of such triangles modulo rotation is just one point.  Next, note
that when $A_{12}$ is neither a minimum or maximum, its level sets give
circles in the leaf, parametrised by the angle $\theta$ between the
plane spanned by $E_1$ and $E_2$ and the plane spanned by $E_3$ and
$E_4$.  (See Fig.\ \ref{circle}.)  It follows that the leaf is a 2-sphere.  

On this 2-sphere, the meridians $\theta = 0$ and $\theta = \pi$ 
correspond to quadrilaterals that lie in a plane in $\R^3$.   In terms
of tetrahedra, points on these meridians correspond to degenerate
tetrahedra of zero volume, for which $U = 0$.  The two hemispheres $0<
\theta< \pi$ and $\pi < \theta < 2\pi$ correspond to the regions where
$U > 0$ and $U < 0$.  Which hemisphere corresponds to which sign of $U$
is a matter of convention in the definition of $\theta$.  We can fix
this convention by demanding that the Hamiltonian vector field generated
by $A_{12}$ be  $\partial_\theta$.  The reason we can demand this is
that $A_{12}$ generates rotations of the vectors $E_1$ and $E_2$ about
the axis  $E_1 + E_2$, while leaving $E_3$ and $E_4$ fixed.  With this
convention $A_{12}$ and $\theta$ are canonically conjugate variables, so
the symplectic structure on the 2-sphere is $\omega = \xd A_{12}\wedge
\xd \theta$.   

\subsection{Quantization} \label{3dtet.quant}

There are two strategies for constructing the space of states of the
quantum tetrahedron in 3 dimensions.   The first is to geometrically quantize
$(\so(3)^\ast)^4$ and impose the closure constraint at the quantum
level.    The second is to impose the closure constraint at the classical
level and geometrically quantize the resulting reduced space $T$.   If the
$\SO(3)$ action generated by the closure constraint were free, we could
carry out both strategies and use a result of Guillemin and Sternberg
\cite{GS2} to show that they give naturally isomorphic vector spaces. 
In short, quantization would commute with reduction.  

Unfortunately, as we saw in previous section, the $\SO(3)$ action
generated by the closure constraint is not free.  This complicates the
second strategy.  Generically $\SO(3)$ acts freely on the symplectic
leaves of $(\so(3)^*)^4$, but there are also leaves on which  the action
is not free.  In these nongeneric cases, the corresponding reduced leaf
in $T$ has singular points where it is not a K\"ahler manifold.   This
presents an obstacle to geometric quantization.   

With enough cleverness one could probably overcome this obstacle and
generalize Guillemin and Sternberg's result to cover this situation.  
Instead, we take a less ambitious approach.   We carry out the first
quantization strategy completely, carry out the second one for generic
leaves of $(\so(3)^*)^4$, and show that the two strategies give
naturally isomorphic vector spaces in the generic case.  

In the first approach we start by geometrically quantizing
$(\so(3)^\ast)^4$.  Since geometric quantization takes products to
tensor products, we obtain the Hilbert space $\H^{\tensor 4}$, where
$\H$ is the Hilbert space of a quantum bivector in 3 dimensions, as
defined in Section \ref{bivect.quant}.  Quantizing the coordinate functions
$E_1^i,\dots,E_4^i$ on $(\so(3)^\ast)^4$, we obtain operators
$\hat{E}_1^i, \dots , \hat{E}_4^i$ on $\H^{\tensor 4}$.   Explicitly, we
have 
\ban \hat E_1^i = \hat{E}^i \tensor 1 \tensor 1 \tensor 1  \\
     \hat E_2^i = 1 \tensor \hat{E}^i \tensor 1 \tensor 1  \\
     \hat E_3^i = 1 \tensor 1 \tensor \hat{E}^i \tensor 1  \\
     \hat E_4^i = 1 \tensor 1 \tensor 1 \tensor \hat{E}^i   \ean
where $i = 1,2,3$.   We then impose the closure constraint at the
quantum level.  This gives us the subspace of $\H^{\tensor 4}$
consisting of states $\psi$ with 
\[      (\hat E_1^i + \hat E_2^i + \hat E_3^i + \hat E_4^i)\psi = 0 .\] 
Such states are precisely those that are invariant under the action of
$\SU(2)$.  We call this subspace the {\it Hilbert space of the quantum 
tetrahedron in 3 dimensions}, and denote it by $\T$.   Thus we have 
\[          \T = \Inv(\H^{\tensor 4}) \cong 
\bigoplus_{j_1,\dots,j_4} \Inv(j_1 \tensor \cdots \tensor j_4)  \]
where `Inv' denotes the invariant subspace.

In the second approach, we start by noting that each integral symplectic
leaf of $(\so(3)^\ast)^4$ is a product of 4 integral symplectic leaves
of $\so(3)^\ast$.  Thus it is of the form
\[   \Lambda = \{  |E_i| = j_i \} \]  
for some spins $j_1,\dots,j_4$.  As a product of K\"ahler manifolds,
it acquires the product K\"ahler structure.  The closure constraint
generates an $\SO(3)$ action which preserves this K\"ahler structure.  
As long as the spins $j_i$ do not satisfy $j_1 + j_2 = j_3 + j_4$, $j_1
= j_2 + j_3 + j_4$ or any permutations of these, this action is free. 
In this case, the reduced leaf $(\Lambda \cap C)/\SO(3)$ becomes a K\"ahler
manifold as well.   Geometrically quantizing this reduced leaf, we obtain 
a space of states.  By the result of Guillemin and Sternberg, this vector
space is naturally isomorphic to the space 
\[           \Inv(j_1 \tensor \cdots \tensor j_4) \subset \T. \]

We can also see this last fact directly without using much machinery.   
The map from invariant states on $\Lambda$ to the state space for 
$(\Lambda \cap C)/\SO(3)$ is an injection.  But we can see that these 
spaces have the same dimension, as follows.
The dimension of the space of invariant states on $\Lambda$ is 
\[    \dim(\Inv(j_1 \tensor \cdots \tensor j_4) = 
\min \{ |j_1 + j_2|, |j_3 + j_4| \} - 
\max \{ \left||j_1| - |j_2|\right|, \left||j_3| - |j_4|\right| \} + 1 \]
On the other hand, the Riemann-Roch theorem implies that
the dimension of the state space for $(\Lambda \cap
C)/\SO(3)$ is $1\over 2\pi$ times its symplectic volume plus 1.  The
symplectic volume was computed by Hausmann and Knutson \cite{HK} to be
$2\pi(A_{\rm max} - A_{\rm min})$, where $A_{\rm min}$ and $A_{\rm max}$
are the minimum and maximum values of $A_{12}$.   This is easy to see
for generic leaves using the results of the previous section:
\[  \int_{(\Lambda \cap C)/\SO(3)} \omega = 
\int_{A_{\rm min}}^{A_{\rm max}} \int_0^{2\pi} \xd A_{12}\wedge \xd \theta =
2\pi(A_{\rm max} - A_{\rm min})   \]
The dimension of the state space for $(\Lambda \cap C)/\SO(3)$ is thus
$A_{\rm max} - A_{\rm min} + 1$.  But since $A_{12} = |E_1 + E_2| = 
|E_3 + E_4|$, we have
\[       A_{\rm min} = \max \{ \left||j_1| - |j_2|\right|, 
\left||j_3| - |j_4|\right| \} , \]
\[       A_{\rm max} = \min \{ |j_1 + j_2|, |j_3 + j_4| \}   .\]
Thus the two dimensions are equal.  

The phase space discussed so far includes both genuine tetrahedra and 
their negatives, which do not satisfy the positivity constraint $U>0$. 
In a slightly different context, the implications of this issue for 
quantum gravity were addressed by Loll \cite{Loll}.  More recently,
Barbieri \cite{Barb} studied the obvious quantization of $U$, 
the Hermitian operator
\[ \hat U= \sum_{i,j,k} 
\epsilon_{ijk}\hat E_1^i \hat E_2^j \hat E_3^k, \]
He showed that if $u$ is an eigenvalue, so is $-u$. This arises because 
the natural antilinear structure map $P$ for representations of $\SU(2)$ 
acts on the quantum state space $\T$, and obeys the relations $P\hat U=
-\hat UP$, $P^2=1$.  Thus $P$ is a symmetry which interchanges eigenvectors 
of $U$ with opposite eigenvalues.  

\section{The Quantum Tetrahedron in 4 Dimensions} \label{4dtet}

We now turn to the 4-dimensional case.  As already noted, a
tetrahedron in $\R^4$ with labelled vertices, modulo translations, is
determined by 3 vectors $e_1, e_2, e_3$.   Associating bivectors
to the faces of the tetrahedron in the usual way:
\[ E_1 = e_2 \we e_3, \; E_2 = e_3 \we e_1, \; E_3 = e_1 \we e_2 ,\]
\[ E_4 = -E_1 - E_2 - E_3, \]
we obtain a 4-tuple of simple bivectors $E_1,\dots, E_4$ that sum to
zero, with all their pairwise sums $E_i + E_j$ also being simple.  The
bivectors $E_i$ are simple because they come from triangles.  Their
pairwise sums are simple because these triangles lie in planes that
pairwise span 3-dimensional subspaces of $\R^4$ --- or equivalently,
that intersect pairwise in lines.  

As before, the map from the vectors $e_i$ to the bivectors $E_i$ is
generically two-to-one: when the vectors $e_i$ are linearly independent,
the $E_i$ uniquely determine the vectors $e_i$ up to sign.  However, not
every collection of simple bivectors $E_1,\dots, E_4$ summing to zero
with simple pairwise sums come from a tetrahedron this way.  Generically
there are 4 possibilities:

\begin{enumerate}
\item The bivectors $E_i$ come from a tetrahedron.
\item The bivectors $-E_i$ come from a tetrahedron.
\item The bivectors $\ast E_i$ come from a tetrahedron.
\item The bivectors $-\!\ast \! E_i$ come from a tetrahedron.
\end{enumerate}

The proof is as follows.  Recall that any nonzero simple bivector $e \we
f$ determines a plane through the origin, namely that spanned by $e$ and
$f$, and note that conversely, this plane determines the simple bivector
up to a scalar multiple.   Generically the bivectors $E_1,E_2,E_3$ are
nonzero, and thus determine three planes $P_1,P_2,P_3 \subset \R^4$. 
Since the pairwise sums of the $E_i$ are simple, these three planes
generically intersect pairwise in lines.  There are two cases:

\begin{alphalist}

\item $P_1 \cap P_2 \ne P_1 \cap P_3$.  In this case $P_1$ is spanned 
by the lines $P_1 \cap P_2$ and $P_1 \cap P_3$.  Thus it lies in the
span of $P_2$ and $P_3$, which is a 3-dimensional subspace of $\R^4$. 
Since all three planes lie in this subspace, the problem reduces to the
3-dimensional case discussed in Section \ref{3dtet}: either the
bivectors $E_i$ come from a tetrahedron, or the bivectors $-E_i$ come
from a tetrahedron.  

\item $P_1 \cap P_2 = P_1 \cap P_3$.  In this case all three planes
$P_i$ share a common line, but generically they pairwise span three
different subspaces, so the span of $P_1$ and $P_2$ differs from that of
$P_1$ and $P_3$.   Taking orthogonal complements, we thus have
$P_1^\perp \cap P_2^\perp \ne P_1^\perp \cap P_2^\perp$.  Thus all the
hypotheses of the previous case hold, but with the planes $P_i$ replaced
by the planes $P_i^\perp$.  The planes $P_i^\perp$ correspond to the
bivectors $\ast E_i$, so the argument in the previous case implies that
either the bivectors $\ast E_i$ come from a tetrahedron, or $-\! \ast \!
E_i$ come from a tetrahedron.  \end{alphalist}

This proof makes no use of the bivector $E_4$, but of course the
situation is perfectly symmetrical.  If one forms planes corresponding
to all 4 bivectors $E_1,\dots,E_4$, one can check that in case (a),
all 4 planes lie in the same 3-dimensional subspace of $\R^4$, while
in case (b), all 4 planes contain the same line.  The two cases are
dual to each other in the sense of projective geometry.

One can easily distinguish between the cases 1-4 listed above using the
self-dual and anti-self-dual parts $E_i^\pm$ of the bivectors $E_i$. 
Thinking of these bivectors as elements of $\so(4)$, we can define the
quantities
\[         U^\pm = \pm \langle E_1^\pm, [E_2^\pm ,E_3^\pm] \rangle ,\] 
using the Killing form and Lie bracket in $\so(4)$.   (Just as in
3 dimensions, we can equivalently define these quantities using any
3 of the bivectors $E_1,\dots,E_4$.)  Then cases 1-4 correspond 
to the following 4 cases, respectively:
\begin{enumerate}
\item $U^+ > 0, U^- < 0$. 
\item $U^+ < 0, U^- > 0$. 
\item $U^+ > 0, U^- > 0$.
\item $U^+ < 0, U^- < 0$. 
\end{enumerate}
Moreover, in all of these cases we have $|U^+| = |U^-|$.  To prove
this, we consider the 4 cases in turn:

\begin{enumerate}
\item If the bivectors $E_i$ come from a tetrahedron, there is a
3-dimensional subspace $V \subset \R^4$ such that all the $E_i$ lie in
$\so(V) \subset \so(4)$.  The results of Section \ref{3dtet} thus imply
that $\langle E_1, [E_2,E_3] \rangle > 0$.  Note that we can choose a
Lie algebra isomorphism $\alpha \maps \so(4) \to \so(3) \oplus \so(3)$
mapping $\so(V)$ to the `diagonal' subalgebra consisting of elements of
the form $(x,x)$.  We can also choose $\alpha$ so that self-dual
elements of $\so(4)$ are mapped to the first $\so(3)$ summand, while
anti-self-dual elements are mapped to the second.  Writing $\alpha(E_i)
= (x_i,x_i)$, we thus have
\[  \langle E_1, [E_2,E_3]\rangle =
\langle (x_1,x_1), [(x_2,x_2),(x_3,x_3)]\rangle = 
2 \langle x_1,[x_2,x_3] \rangle \]
while 
\[  U^+ = \langle (x_1,0), [(x_2,0),(x_3,0)]\rangle =
\langle x_1,[x_2,x_3] \rangle \]
and
\[  U^- = -\langle (0,x_1), [(0,x_2),(0,x_3)]\rangle =  
-\langle x_1,[x_2,x_3] \rangle .\]
It follows that $U^+ > 0$ and $U^- = -U^+$.
\item If the bivectors $-E_i$ come from a tetrahedron, since multiplying
the $E_i$ by $-1$ reverses the sign of both $U^+$ and $U^-$, the above
argument implies that $U^+ < 0$ and $U^- = -U^+$.
\item If the bivectors $\ast E_i$ come from a tetrahedron, since applying
the Hodge star operator to the $E_i$ reverses the sign of 
$U^-$ but leaves $U^+$ unchanged, the above argument
implies that $U^+ > 0$ and $U^- = U^+$.
\item Similarly, if the bivectors $-\! \ast \! E_i$ come from a tetrahedron,
we have $U^- < 0$ and $U^- = U^+$.
\end{enumerate}  
The remaining, non-generic, cases correspond to situations where both
$U^+$ and $U^-$ vanish.  

We may visualize the results of this section as follows.  Suppose the
bivectors $E_1,\dots,E_4$ come from a nondegenerate tetrahedron $\tau$ in 4
dimensions.  Then the bivectors $E_i^+$ come from a unique nondegenerate
tetrahedron $\tau^+$ with positive oriented volume $\sqrt{U^+}$ in the 
3-dimensional space of self-dual bivectors.  Similarly, the $E_i^-$ come 
from a unique nondegenerate tetrahedron $\tau^-$ with positive oriented 
volume $\sqrt{-U^-}$ in the space of anti-self-dual bivectors.  
Since $|E_i^+| = |E_i^-|$ and $|E_i^+ + E_j^+| = |E_i^- + E_j^-|$, the 
tetrahedra $\tau^+$ and $\tau^-$ differ only by a rotation.  Conversely, 
any pair of tetrahedra $\tau^+$, $\tau^-$ with these properties 
determines a nondegenerate tetrahedron $\tau$ in 4 dimensions, unique 
up to the transformation $x \mapsto -x$.

\subsection{Poisson structure} \label{4dtet.poisson}

We obtain the phase space of the tetrahedron in 4 dimensions by starting
with phase space of 4-tuples of bivectors and then imposing suitable
constraints.  The phase space of 4-tuples of bivectors is the product of
4 copies of $\so(4)^\ast$ with the `flipped' Poisson structure described
in Section \ref{bivect.poisson}.  This space has coordinate
functions $E^{\pm j}_1, \dots , E^{\pm j}_4$ where $j = 1,2,3$,
satisfying the Poisson bracket relations:
\ban  
      \{ E^{+j}_i, E^{+k}_i \} &=&  \epsilon^{jkl} E^{+l}_i \\
      \{ E^{-j}_i, E^{-k}_i \} &=& -\epsilon^{jkl} E^{-l}_i \\
      \{ E^{+j}_i, E^{-k}_i \} &=&  0 \\
      \{ E^{\pm j}_i, E^{\pm k}_{i'} \} &=&  0 \qquad i \ne i'  .
\ean

To obtain the phase space of a tetrahedron in 4 dimensions, we 
do Poisson reduction using the following constraints:
\begin{enumerate}
\item The closure constraint $E_1 + E_2+ E_3 + E_4 = 0$.
\item The simplicity constraints $|E_i^+|^2 = |E_i^-|^2$.
\item The simplicity constraints $|E_i^+ + E_j^+|^2 = |E_i^- + E_j^-|^2$.
\item The chirality constraint $U^+ + U^- = 0$.  
\end{enumerate}
The last constraint eliminates the fake tetrahedra of types 3 and 4, 
leaving genuine tetrahedra (type 1) and their negatives (type 2).

First we deal with the closure constraint.  This is equivalent to
separate closure constaints for the self-dual and anti-self-dual parts
of the $E_i$:
\[  E_1^+ + \cdots + E_4^+ = 0, \qquad  E_1^- + \cdots + E_4^- = 0 \]
Thus Poisson reduction by this constraint takes us from the
24-dimensional space $(\so(4)^\ast)^4$ down to
12-dimensional space $T^+ \times T^-$, where $T^+$ is a copy of the
phase space for a tetrahedron in 3 dimensions, and $T^-$ is another
copy, but with minus the standard Poisson structure.  

By the results of Section \ref{3dtet.poisson}, the functions $|E_i^\pm|^2$ 
push down to smooth functions on $T^+ \times T^-$, and the symplectic leaves
of $T^+ \times T^-$ are the sets on which all 8 of these functions are
constant.   Generically these leaves are of the form $S^2 \times S^2$.  
The following functions also push down to smooth functions on $T^+
\times T^-$:
\[      C_i = |E_i^+|^2 - |E_i^-|^2 ,\]
\[     C_{ij} = |E_i^+ + E_j^+|^2 - |E_i^- + E_j^-|^2  .\]
The $C_i$ have vanishing Poisson brackets with every function on $T^+
\times T^-$.  The $C_{ij}$ do not, and in particular, taking $ijkl$ to
be any even permutation of $1234$, we have
\[     \{C_{ij},C_{ik}\} = 4(U^+ + U^-). \]

Next we deal with the simplicity constraints $C_i = 0$.  Since the $C_i$
have vanishing Poisson brackets with every function, they generate trivial 
flows, so in this case Poisson
reduction merely picks out the 8-dimensional subspace
\[    \{C_i = 0 \} \subset T^+ \times T^- .             \]
A point $(t^+,t^-)$ of $T^+ \times T^-$ corresponds to a pair of tetrahedra or
negative tetrahedra modulo rotations in 3 dimensions.  When we refer to 
$t^+$, for example, as a tetrahedron, this means the geometrical tetrahedron 
formed either by the vectors $t^+$ if $U^+ \ge 0$ or that formed by $-t^+$ 
if $U^+ \le0$.

A point in the subspace $\{C_i = 0\}$
corresponds to such a pair whose corresponding faces have equal
areas.   The symplectic leaves $L$ of this subspace are again 
generically of the form $S^2 \times S^2$, and consist of the pairs of 
tetrahedra in 3 dimensions for which the common values for their four 
face areas are constants.  Recall that each $S^2$ is divided into two 
hemispheres on which $U \ge 0$ and $U \le 0$. The equator $\{U=0\}$ is 
the circle of degenerate tetrahedra.

Now we consider the simplicity constraints $C_{ij} = 0$ and
the chirality constraint $U^+ + U^- = 0$.  On the subspace
$\{C_i = 0\}$ only two of the constraints $C_{ij} = 0$ are independent, 
since for any permutation $ijkl$ of $1234$ we have
\[         C_{ij} = C_{kl}  ,\]
and we also have
\[         C_{12} + C_{23} + C_{31} = C_1 + C_2 + C_3 + C_4 . \]
Thus the subspace 
\[    \{C_i = C_{ij} = 0 \} \subset T^+ \times T^-              \]
is generically 6-dimensional.  To describe its structure, it is easiest to 
consider its intersection with a particular symplectic leaf $L$. 

The constraints $C_{ij}=0$ imply that the two tetrahedra $t^+,t^-$ are 
isometric. Then $t^+=t^-$ or $t^+=-{t}^-$, and these are both true 
if and only if $U^+ = U^- = 0$.  This means that the constraints are 
satisfied on two copies of $S^2$ embedded in $L$ which intersect 
in the circle $\{U^+ = U^- = 0\}$.  These embedded spheres, which we 
denote by $X$ and $\bar X$, satisfy the equations $U^+ + U^- = 0$ and 
$U^+ - U^- = 0$ respectively. 

The surfaces $X$ and $\bar X$ differ markedly with respect to the symplectic 
structure. It is worth comparing the situation with the K\"ahler reduction 
by a group action.  In the latter, the reduced phase space is the space of 
orbits of the group action in the constraint surface. The constraint surface 
is coisotropic, i.e., the tangent space contains its symplectic complement, 
and the tangents to the orbits are in this symplectic complement. This means 
that if $v$ is tangent to the orbit and $u$ is tangent to the constraint 
surface, then $\omega(v,u)=0$. The wavefunctions which are invariant under 
the group action localise to a small region around the constraint surface 
and are determined by their values on it.

The situation we have here with the constraints $C_{ij}=0$ differs in that 
they do not form a Lie algebra.  However $X$ is coisotropic (in fact 
Lagrangian) and the fact that the commutator of two of the $C_{ij}$ 
vanishes on $X$ implies that the Hamiltonian vector fields they generate 
are tangent to $X$.  Thus the situation is in most respects similar to 
the group case. On the other hand, $\bar X$ is not coisotropic, and the 
Hamiltonian vector fields are not tangent to it.  Thus there is no 
sensible reduction procedure for wavefunctions based on $\bar X$. 

\subsection{Quantization} \label{4dtet.quant}

Finally, let us quantize the tetrahedron in 4 dimensions.
As in 3 dimensions, there are two strategies.  The first is 
to geometrically quantize $(\so(4)^\ast)^4$ and impose the 
closure, simplicity and chirality constraints at the quantum
level.    The second is to impose these constraints at the classical
level and geometrically quantize the resulting reduced space.
As we have already seen, there is a problem with the second approach:
the simplicity constraints $C_{ij}$ do not generate a Lie group
action on the symplectic leaf $L$.  Thus we can only carry out the second 
strategy in a rather ad hoc way.  If we do so, we obtain a 1-dimensional
space of states for each generic integral leaf $L$.  In some sense this 
explains the uniqueness of the vertex.   Unfortunately, we cannot apply 
the results of Guillemin and Sternberg to {\it rigorously} conclude that 
the first strategy also gives a 1-dimensional state space for a tetrahedron 
with faces of fixed area.  

In the first strategy, we start by geometrically quantizing the product
of four copies of $\so(4)^\ast$ with its flipped Poisson structure.  
By the results of Section \ref{bivect.quant} we obtain the Hilbert space
$(\H \tensor \H^\ast)^{\tensor 4}$, together with operators
$\hat E_i^{\pm j}$ on this space which satisfy the following commutation 
relations:
\ban
\lbrack {\hat E_i}^{+j}, {\hat E_i}^{+k} \rbrack 
&=& i\epsilon^{jkl} {\hat E_i}^{+l}  \\
\lbrack {\hat E_i}^{-j}, {\hat E_i}^{-k} \rbrack 
&=& -i\epsilon^{jkl} {\hat E_i}^{-l}  \\
\lbrack {\hat E_i}^{+j}, {\hat E_i}^{-k} \rbrack &=&  0  \\
\lbrack {\hat E_i}^{\pm j}, {\hat E_{i'}}^{\pm k} \rbrack &=&  0 
\qquad i \ne i'. 
\ean
Geometrically quantizing the closure constraint we obtain the
operators 
\[                \hat E_1^\pm  + \cdots + \hat E_4^\pm , \]
while quantizing the simplicity constraints gives the operators
\[     \hat C_i = \sum_{k=1}^3 (\hat E^k_i)^2 , \qquad
 \hat C_{ij} = \sum_{k=1}^3 (\hat E^k_i+ \hat E^k_j)^2 . \]
The states annihilated by the quantized closure constraint form the subspace
\[        \T \tensor \T^\ast  \subset (\H \tensor \H^\ast)^{\tensor 4} \]
where $\T \subset \H^{\tensor 4}$ is the Hilbert space of the quantum
tetrahedron in 3 dimensions.   Among these states, those that are also
annihilated by the simplicity constraints $\hat C_i$ form the subspace
\[  \bigoplus_{j_1,j_2,j_3,j_4} 
\Inv((j_1,j_1)\otimes \cdots \otimes(j_4,j_4))  \subset \T \tensor \T^* .\]
The operators $\hat C_{ij}$ map each of these summands to itself, and as
shown by Reisenberger \cite{R2}, each nonzero summand has a 1-dimensional 
subspace that is annihilated by all these operators.  We call the direct
sum of all these 1-dimensional spaces {\it the Hilbert space of the
quantum tetrahedron in four dimensions}.

The reader may wonder why we have not dealt with the chirality 
constraint.  If we geometrically quantize this constraint we 
obtain the operator $\hat U^+ + \hat U^-$, where 
\[ \hat U^{\pm} = \sum_{i,j,k}
\epsilon_{ijk}\, \hat E_1^{\pm i} \,\hat E_2^{\pm j} \,\hat E_3^{\pm k}. \]
Barbieri \cite{Barb2} has shown that
\[       4i [ \hat C_{ij}, \hat C_{ik} ] = \hat U^+ + \hat U^-. \]
Thus any solution $\psi$ of the simplicity constraints 
automatically satisfies the chirality constraint 
\[        (\hat U^+ + \hat U^-)\psi = 0 .\]
In other words, fake tetrahedra of types 3 and 4 do not 
occur in the quantum theory.  

In the second strategy, we attempt to impose all the constraints at 
the classical level and then quantize.  
We did most of the work for this in the previous section.  Starting
from $(\so(4)^\ast)^4$ and doing symplectic reduction using the 
closure constraint, we obtained the space $T^+ \times T^-$.  Imposing
the simplicity constraints $C_i = 0$ we obtained an 8-dimensional 
space with symplectic leaves $L$ generically of the form 
$S^2 \times S^2$.   Recall that any point in $L$ corresponds to a pair
$(t^+,t^-)$ of tetrahedra in 3 dimensions (modulo rotations) such that 
the $i$th face of both $t^+$ and $t^-$ has area $j_i$.  If at this 
stage we geometrically quantized an integral generic leaf $L$, 
we would obtain the Hilbert space of states
\[        \Inv((j_1,j_1)\otimes \cdots \otimes(j_4,j_4)),  \]
which is one of the summands mentioned above.  We could then 
impose the remaining simplicity constraints at the quantum level as
before.  However, let us impose the simplicity constraints $C_{ij} = 0$ 
at the classical level.   We obtain a union of two spheres $X \cup \bar X$.  
The chirality constraint holds only on $X$, so we restrict attention to 
this space.

As already noted, the constraints $C_{ij}$ do not generate a Lie group
action on $L$.  However, they generate flows that act transitively on the 
sphere $X$, in the sense that all smooth functions invariant under all these 
flows are constant on $X$.  Thus we may say, in a somewhat ad hoc way, 
that the symplectic reduction of the leaf $L$ by the constraints $C_{ij}$ 
is a single point.  Geometrically quantizing this point, we obtain a 
1-dimensional space of states, in accord with Reisenberger's result.  
Alternatively, we may think of this 1-dimensional space as the space of
constant functions on $X$.  This makes the uniqueness of the vertex much
less mysterious.

Unfortunately, we cannot use this to give a new proof of Reisenberger's 
result, because we do not know that quantization commutes with reduction 
in this case.  The reason is that the constraints $C_{ij}$ do not generate
a Lie group action on $L$.  Moreover, the flows they generate do not
preserve the K\"ahler structure on $L$.  A complete proof of the uniqueness
of the vertex using geometrical quantization will apparently require
new ideas. Flude has studied some similar constraints which give results in this direction which hold in the asymptotic limit of large spin \cite{Flude}.

\section{Considerations from Quantum Gravity} \label{considerations}

The action for Riemannian general relativity in 4 dimensions, expressed
in terms of a linear connection and an $\R^4$-valued one-form $e$, can be 
written as:
\[  \int \langle F, * (e\wedge e)\rangle  \]
where $F$ is the $\so(4)$-valued 2-form giving the curvature, $e\wedge e$ 
is the induced bivector-valued 2-form, and $*$ the Hodge star. The brackets
$\langle\cdot\,,\cdot\rangle$ denote the standard pairing of $\so(4)$ with 
$\Lambda^2 \R^4$ introduced in section \ref{bivect}, extended to differential 
forms by using the exterior product of forms.  The inner product 
$\langle\cdot\,,\ast\,\cdot\rangle$ is exactly the one that leads to the 
flipped Poisson structure we introduced.  This means that the Poisson 
structure we introduced so that the quantization leads to real tetrahedra, 
and not fake tetrahedra, is exactly the one relevant for the quantization 
of the Einstein action.

A further issue is the negative tetrahedra, the ones of type 2 for which the 
bivectors $-E$ come from a tetrahedron. These configurations make up half of 
the phase space $X$. There is a classical symmetry $E\mapsto -E$, and a 
corresponding quantum operator, $P\otimes P$, with $P$ the antilinear operator 
discussed in Section \ref{3dtet.quant}. Since this commutes with the quantum 
constraints, $(P\otimes P)\psi = c\psi$, with $c$ a complex number of unit 
modulus. The interpretation of this is that the quantum tetrahedron in four 
dimensions contains both tetrahedra and negative tetrahedra in equal measure.  
It does not seem that one can easily exclude the negative tetrahedra.  There 
is a natural interpretation which may provide an explanation.  One can 
change the sign of the isomorphism $\beta\colon\Lambda^2 \R^n\to\so(n)^\ast$ 
by replacing the Euclidean metric $\eta$ by $-\eta$. Thus the negative 
tetrahedra have an interpretation as genuine tetrahedra for the opposite 
signature of the space-time metric.  

Finally, there remains the question of the implication of our results to the 
nature of quantum geometry.  An approach based on areas is in many ways a 
natural extension of 3-dimensional quantum gravity based on lengths.  However 
it differs rather substantially from a naive expectation of a 4-dimensional 
approach based on lengths.  As we have seen, a quantum tetrahedron does not 
appear to have a unique metric geometry.  This means that in a 
state-sum approach based on gluing 4-simplexes together across a common 
tetrahedron, as outlined in \cite{BC}, the metric of the tetrahedron does not 
transmit across from one 4-simplex to another.  Rather, the parallelogram 
areas are randomised as one crosses from one 4-simplex to another.  As we 
have shown here, this is a direct consequence of the uncertainty principle. 

\subsection*{Acknowledgments}

We would like to thank Andrea Barbieri, Laurent Freidel, Allen Knutson,
Kirill Krasnov, Michael Reisenberger, Alan Weinstein, and Jos\'e-Antonio
Zapata for useful discussions and correspondence.  In particular, we
thank Zapata for pointing out the importance of the flipped Poisson
structure.

\end{document}